\documentclass[vecphys]{svmult}

\usepackage{makeidx}         % allows index generation
\usepackage{graphicx}        % standard LaTeX graphics tool
\usepackage{multicol}        % used for the two-column index
\usepackage[bottom]{footmisc}% places footnotes at page bottom
\makeindex             % used for the subject index
\newcommand{\simgeq}{{\widetilde{>}}}
\begin{document}

\title*{FIGGS: Faint Irregular Galaxies GMRT Survey}
\author{Ayesha Begum\inst{1}\and
Jayaram N. Chengalur\inst{2}\and
Igor D. Karachentsev\inst{3}\and 
Margrita Sharina\inst{3}\and
Serafim Kaisin\inst{3}
}
\institute{Institute of Astronomy, University of Cambridge, Madingley Road, Cambridge
\texttt{ayesha@ast.cam.ac.uk}
\and National Centre for Radio Astrophysics, TIFR, Pune University Campus, Ganeshkhind, Pune
\and Special Astrophysical Observatory, Nizhnii Arkhys 369167, Russia}
\authorrunning{Begum et al.} 
\maketitle

\section{Introduction}
\label{sec:1}
% Always give a unique label
% and use \ref{<label>} for cross-references
% and \cite{<label>} for bibliographic references
% use \sectionmark{}
% to alter or adjust the section heading in the running head

     HI 21cm aperture synthesis observations of spiral galaxies is a mature
field with over two decades of history -- probably something of the order
of a thousand galaxies have already been imaged. However, the observations
have tended to focus on bright ($\sim$ L$_*$) galaxies with HI masses
$\sim 10^9$~M$_\odot$. Dwarf galaxies (M$_B \simgeq -17$) require substantial
investments of telescope time, and have hence not been studied in similar
numbers. 

To start addressing this imbalance, we have been conducting an HI imaging study of
 faint dwarf galaxies $-$ the Faint Irregular Galaxies GMRT Survey (FIGGS).  The immediate
goal of FIGGS is to obtain high quality observations of the atomic ISM in a large,
volume limited sample of faint, gas rich, dwarf irregular galaxies. Here we briefly describe
the survey and discuss some of the science that we anticipate can be done with this 
data set.

\subsection{FIGGS Sample}
\label{sec:2}

\begin{figure}
%\centering
\includegraphics[height=6cm]{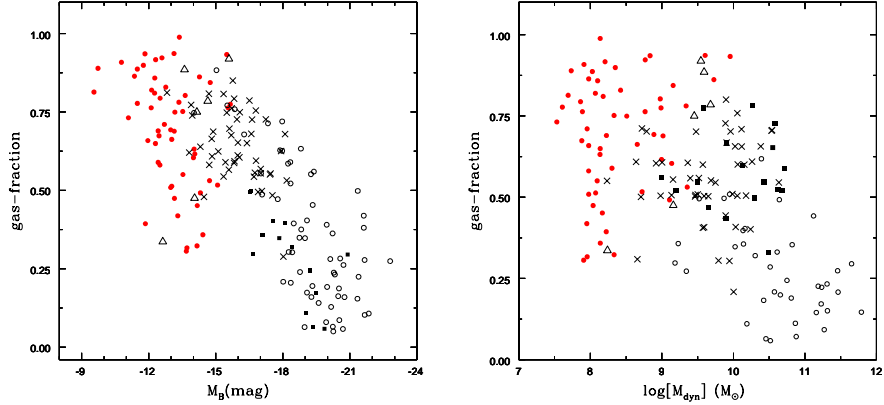}
\caption{
The gas fraction of FIGGS galaxies (red points) and previously
studied galaxies (black points) plotted as functions of
absolute blue magnitude (left) and dynamical mass (right).
Note how the GMRT FIGGS sample extends the coverage of all three
galaxy properties.}
\label{fig:ayesha_gasfract}       % Give a unique label
\end{figure}

The FIGGS galaxies form an HI flux limited subsample of the Karachentsev et al.(2004) catalog 
of galaxies within 10 Mpc.  Specifically, the FIGGS sample consists of  65 faint dwarf irregular (dIrr) 
galaxies with M$_B \widetilde{>}-14.5$, HI flux integral $\widetilde{>}$ 1 Jy kms$^{-1}$ and
optical sizes $\widetilde{>}$ 1 arcmin. The FIGGS galaxies represent the extreme low-mass end 
of the dIrr population, with a median ${\rm{M_B}} \sim -13$ and a median HI mass 
$\sim 3 \times 10^7$~M$_\odot$. Fig.~\ref{fig:ayesha_gasfract} compares the distributions of
gas fraction, luminosity and dynamical mass of the FIGGS galaxies with that of existing samples
of galaxies with HI aperture synthesis observations. As can be seen, the FIGGS survey
substantially extends the region of parameter space which has largely gone untouched by previous
HI imaging studies.

\begin{figure}
%\centering
\includegraphics[height=3cm]{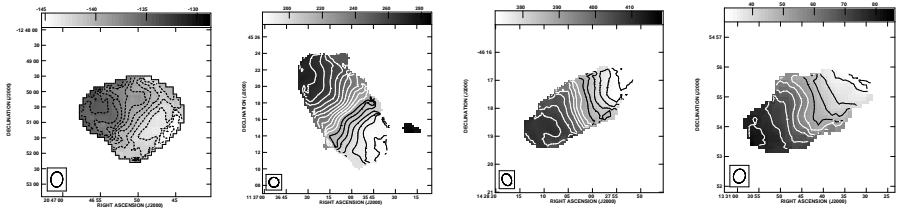}
\caption{ The regular velocity fields of some of the faintest FIGGS galaxies.
}
\label{fig:ayesha_velfield}       % Give a unique label
\end{figure}

The typical GMRT integration time per source for most galaxies is $\sim 5 - 6$
hours, which gives a typical rms of $\sim$ 2-3 mJy/Beam per channel. It is worth 
emphasising that our observations used a relatively high velocity resolution 
($\sim$1.6 kms$^{-1}$, i.e. $\sim$ 4 times better than most earlier interferometric
studies of such faint dwarf galaxies). This high velocity resolution is crucial to 
detect large scale velocity gradients in the faintest dwarf galaxies. Our observations
show that, (unlike what one is lead to believe from coarser velocity resolution
observations, e.g.  Lo et al. 1993), most faint dwarf irregular galaxies in fact
have large scale systematic patterns in their velocity fields (Fig.~\ref{fig:ayesha_velfield},
see also Begum et. al. 2006). Galaxies from the FIGGS sample are the faintest known galaxies 
to show such regular kinematics. 

The GMRT HI images are supplemented by single dish HI observations,  HST  V and I band imaging 
of the resolved stars and  ground based H$\alpha$ imaging using 6-m BTA telescope. Distances 
accurate to $\sim$ 10\% are available for most of the galaxies in our sample $-$ the FIGGS sample 
is the first large sample of faint dIrr for which interferometric data is available and 
distances are accurately known. Additionally,  the HII region abundances and H$\alpha$ rotation 
curves are being obtained on the WHT, INT telescopes on La Palma and 6-m BTA telescope respectively.

\subsubsection{Science Drivers for FIGGS }

\begin{figure}
\centering
\includegraphics[height=3.5cm]{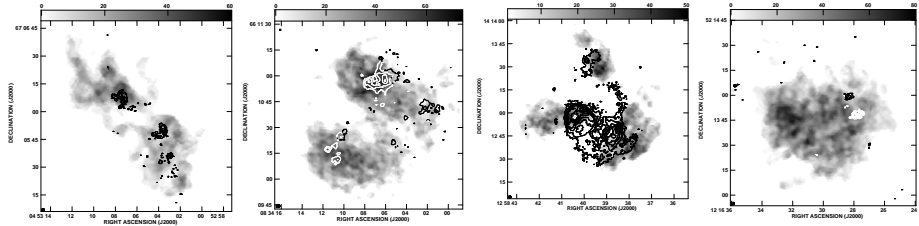}
\caption{
GMRT integrated HI images of some of FIGGS galaxies
(greyscales) at 3$^{\prime\prime}$ resolution (corresponding to a linear
scale of $\sim$ 30-90 pc) overlayed on H$\alpha$ images (contours).
}
\label{fig:ayesha_highres}       % Give a unique label
\end{figure}

One of the main goals of FIGGS is to use the HI images in conjugation with the 
optical data to study the interplay between the neutral ISM and star formation in the faintest,
lowest mass, gas rich dIrr galaxies. The FIGGS data will enable us to study the ISM of most 
of our sample galaxies at a linear resolution of $\sim15-150$ pc $-$ i.e. comparable to the  scales
at which energy is injected into the ISM through supernova and stellar winds. FIGGS thus  provide
a unique opportunity to study the feedback of star formation in low mass, gas rich galaxies,
 which in turn will allow us to understand the processes driving star formation in these galaxies.
Examples of GMRT HI maps at 3$^{\prime\prime}$ resolution (corresponding to linear scale of $\sim$ 30$-$90 pc)
are shown in Fig.~\ref{fig:ayesha_highres}.

The second major aim of this survey is to extend the Baryonic Tully Fisher relation
(Mcgaugh et. al. (2000)) to a regime of very low mass/luminosity that 
has not yet been well explored. While for the brighter galaxies
W$_{50}$ (the velocity width at 50\% emission), once corrected for random motions and
instrumental broadening, is a good measure of the rotational velocity of the galaxy
(Verheijen 1999), this is not true in the case of faint
dwarf galaxies where random motions could be comparable to the peak rotational velocities
(e.g. Begum et al. 2003). For such galaxies,
it is important to accurately correct for the pressure support (``asymmetric drift''
correction) for which one needs to know both the rotation curve as well as the distribution
of the HI gas, both of which can only be obtained by interferometric observations such as
in FIGGS. 

Our final objective is to use the HI kinematics of FIGGS galaxies, in conjunction
with the H$\alpha$ rotation curves to accurately determine the density distribution of 
the dark matter halos of faint galaxies. Serendipitous discoveries are an added
bonus -- for example, FIGGS has discovered some very extended HI disks around dwarf galaxies e.g.
GMRT HI images of NGC 3741 (M$_B\sim-13.0$) showed it to have an HI extent of $\sim$ 8.3 times 
R${\rm{_{Ho}}}$ (Holmberg radius) (follow-up WSRT+DRAO+GMRT observations found
$\sim$8.8 R${\rm{_{Ho}}}$) $-$ NGC 3741 has the most extended HI disks and  we could derive a rotation 
curve upto a record of 38 times the disk scale length. NGC~3741 has M${\rm{_D/L_B}} \sim 107$ $-$ 
which makes it one of the ``darkest'' irregular galaxies known (Begum et al. 2005).

As a service to the community, calibrated (u,v) data, data cubes, MOMNT maps, rotation curves etc. 
from the FIGGS survey will be publicly released at the end of our survey.

%
%
% BibTeX users please use
% \bibliographystyle{}
% \bibliography{}
%
% Non-BibTeX users please follow the syntax
% the syntax of "referenc.tex" for your own citations
%\input{referenc}
%%%%%%%%%%%%%%%%%%%%%%%%%%%%%%%%%%%%%%%%%%%%%%%%%%%%%%%%%%%%%%%%%%%%%%  }

%%%%%%%%%%%%%%%%%%%%%%%%%%%%%%%%%%%%%%%%%%%%%%%%%%%%%%%%%%%%%%%%%%%%%%

\printindex
\end{document}